\documentclass[11pt,onecolemn]{revtex4}

\usepackage{graphics}
\usepackage{amssymb}
\usepackage[english,francais]{babel}

\begin{document}

\selectlanguage{english}
\title{Study of Efimov physics in two nuclear-spin sublevels of $^7$Li}
\author{Noam Gross$^{1}$, Zav Shotan$^{1}$, Olga Machtey$^{1}$, Servaas Kokkelmans$^{2}$ and Lev Khaykovich$^{1}$}
\affiliation{$^{1}$Department of Physics, Bar-Ilan University,
Ramat-Gan, 52900 Israel,}\affiliation{$^{2}$Eindhoven University
of Technology, P.O. Box 513, 5600 MB Eindhoven, The Netherlands}

\begin{abstract}
Efimov physics in two nuclear-spin sublevels of bosonic lithium is
studied and it is shown that the positions and widths of
recombination minima and Efimov resonances are identical for both
states within the experimental errors which indicates that the
short-range physics is nuclear-spin independent. We also find that
the Efimov features are universally related across Feshbach
resonances. These results crucially depend on careful mapping
between the scattering length and the applied magnetic field which
we achieve by characterization of the two broad Feshbach resonances
in the different states by means of rf-spectroscopy of weakly bound
molecules. By fitting the binding energies numerically with a
coupled channels calculation we precisely determine the absolute
positions of the Feshbach resonances and the values of the singlet
and triplet scattering lengths.

\end{abstract}
\maketitle

\selectlanguage{english}
% main text
\section{Introduction}

The recent years' remarkable progress in the study of Efimov quantum
states in ultracold atoms has renewed a great deal of interest in
this "old-new" quantum few-body problem. Since the first prediction
made by V. Efimov in the early 70's~\cite{Efimov}, many systems were
considered for an experimental study of these quantum states,
however all of them were attempted in vain. The first experimental
evidence of Efimov physics was reported in 2006 in a system of
ultracold $^{133}$Cs atoms~\cite{Kraemer06} which was later on
enhanced and verified in an additional study of the same
system~\cite{Knoop09}. Since then, signatures of Efimov physics have
been observed in other ultracold atomic species which turn out to be
the only platform up to now suitable to study Efimov physics.

It is well known that at very low collision energies the only
partial wave contributing to the scattering process is the s-wave

and thus the two-body interaction is completely determined by the
s-wave scattering length $a$. When $a$ exceeds the characteristic
two-body potential range $r_{0}$, weakly bound three-body Efimov
states emerge and their number scales logarithmically with $a$,
$N_{b}=(s_{0}/\pi)\ln(|a|/r_{0})$ where
$s_{0}=1.00624$~\cite{Efimov}. When $|a|\rightarrow\pm \infty$,
the number of bound states goes to infinity and their energies are
related in powers of a universal scaling factor $\exp(-2\pi/s_{0})
\approx 1/(22.7)^2$~\cite{Efimov,Braaten&Hammer06}. A first
indirect evidence of two consecutive Efimov states was
demonstrated in ultracold $^{39}$K~\cite{Zaccanti09} and a
universal scaling across a region of $|a|\rightarrow\pm \infty$
was verified in $^7$Li~\cite{Gross09}.

Studies of Efimov physics have been rapidly extended beyond the spin
polarized bosonic samples. A notable example of a three-fermion
system, all in different spin states, has been shown by a number of
experimental groups~\cite{Ottenstein08,Huckans09,Nakajima10} one of
which developed a new and promising experimental approach to probe
directly the Efimov quantum states~\cite{Lompe10}. Another example
of a different system is heteronuclear universal trimers observed in
Ref.~\cite{Barontini09}.

Recent developments extended the universal few-body physics to the
domain of four-body states which was theoretically
predicted~\cite{Stecher09} and experimentally
verified~\cite{Ferlaino09}. Many three- and four-body features over
a large dynamical range have been demonstrated in $^7$Li in
Ref.~\cite{Pollack09}. Interestingly, the universal few-body physics
can be extended to N-body clusters~\cite{Stecher10} though their
experimental confirmation remains obscure.

In this work we summarize our study of Efimov physics in two
different energy sublevels and across two different Feshbach
resonances of the same atomic system~\cite{Gross09,Gross10}. As both
Feshbach resonances occur at high magnetic fields where nuclear and
electronic spins are decoupled, the two energy sublevels are
associated with two nuclear-spin states. Our main finding is that
the Efimov features are identical within the experimental errors in
these two states. The absolute location and lifetime of an Efimov
state is defined by the unknown short-range part of the three-body
potential. Most generally, the short range potential is given in
terms of two-body potential permutations of the two-body subsystems
and a true three-body potential which is of importance only when
three particles are very close together. It is very difficult to
solve the short-range physics exactly, and therefore this region is
usually treated in terms of a three-body
parameter~\cite{Braaten&Hammer06,Dincao09}. Thus, our results should
be interpreted as a proof that the three-body parameter (and thus
the short range physics) is identical for the two states. We provide
new and accurate characterizations of Feshbach resonances on both
states and reevaluate our previously published results in accordance
with this study. Though small changes in the positions of the Efimov
features can be identified, the main conclusion is not affected.

This paper is organized as follows. Study of Efimov physics across
two different Feshbach resonances in two nuclear-spin states is
reported in Section~\ref{Sec-Efimov}. We start by comparing the
relevant properties of the two energy levels and their resonances.
Then, after a description of the experimental procedure we report on
newly fitted positions and widths of the Efimov features based on
our most recent and precise characterization of the Feshbach
resonances on both nuclear-spin states. The latter is based on
fitting of the binding energies of weakly bound molecular states
,obtained by RF spectroscopy, to a coupled channels calculation
locally for each resonance and globally for both resonances. This
study is crucial for an accurate mapping of the magnetic field to
the scattering length and it is considered in
Section~\ref{Sec-FeshbachResonances}. Finally,
Section~\ref{Sec-Conclusions} concludes the paper.

\section{Efimov physics across Feshbach resonances}
\label{Sec-Efimov}

\subsection{Comparison between two nuclear-spin states}

\begin{figure*}[h]
{\centering \resizebox*{1\textwidth}{0.28\textheight}
{{\includegraphics{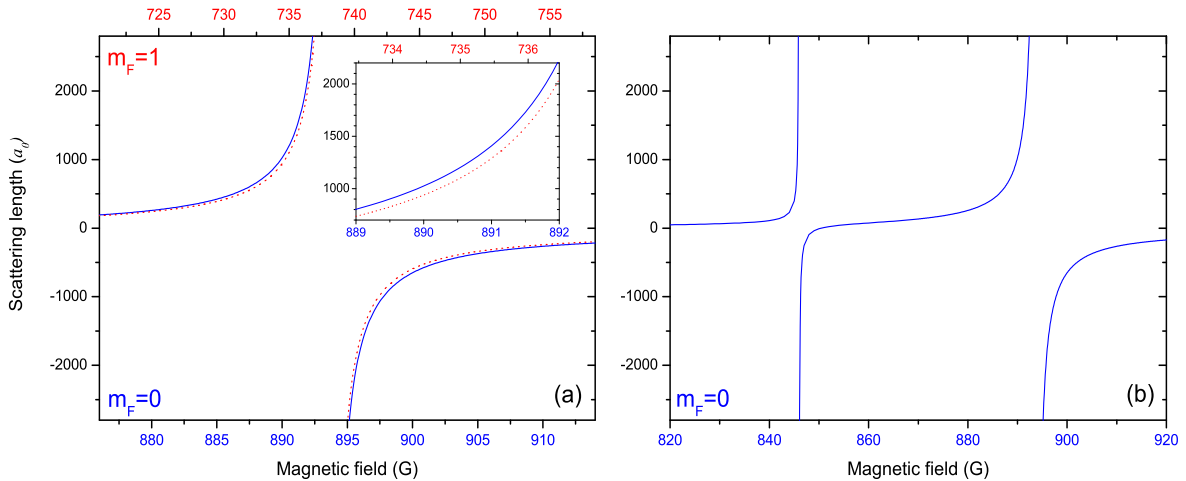}}}
\par}
\caption{\label{Feshbach}(a) The broad Feshbach resonances of the
$|m_{F}=1\rangle$ (dashed red line) and the $|m_{F}=0\rangle$ (solid
blue line) states, centered at $738.2~G$ and $893.7~G$,
respectively. (b) The two Feshbach resonances of the
$|m_{F}=0\rangle$ state.}
\end{figure*}

A comprehensive discussion on the experimental characterization of
Feshbach resonances will be the subject of
Section~\ref{Sec-FeshbachResonances}. Here we intend to draw a
general comparison between the broad resonances of the
$|m_{F}=1\rangle$ and the $|m_{F}=0\rangle$ states of $^7$Li in the
context relevant to the study of Efimov physics. In
Fig.~\ref{Feshbach}(a) we show the scattering length in units of
Bohr radius $a_{0}$ as a function of magnetic field in the vicinity
of both Feshbach resonances with their centers aligned. It can be
easily recognized that the resonances are comparable in widths while
the one on the $|m_{F}=0\rangle$ state is slightly wider. Although
there are different hyperfine states involved, these two resonances
have their origin in the same molecular bound state. Another
difference is that in contrast to the $|m_{F}=1\rangle$ state where
only one resonance exists, on the $|m_{F}=0\rangle$ state there is a
second resonance overlapping with the first one. It is much narrower
and is positioned by atom loss measurement at $\sim 845$~G
\cite{Gross09} as shown in Fig.~\ref{Feshbach}(b). We note that the
measurements reported here are obtained in close vicinity to the
wide resonance's position, away from the narrow one by many times
its width, thus it is not expected to influence the results in the
region of interest.

\begin{figure*}[h]
{\centering \resizebox*{0.6\textwidth}{0.28\textheight}
{{\includegraphics{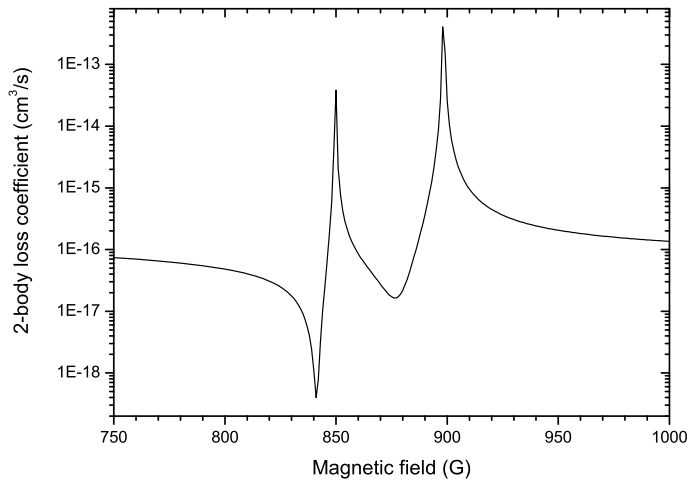}}}
\par}
\caption{\label{2body}A coupled-channels calculation of dipolar
relaxation rate coefficients as a function of magnetic field for
the $|m_{F}=0\rangle$ state.}
\end{figure*}

Signature of Efimov physics is studied here by measuring three-body
recombination loss of atoms as it has been investigated in all but
one recent experiments. In that respect, there is an inherent
difference between the two states: in the absolute ground state,
two-body relaxation mechanism is fundamentally forbidden while in
the $|m_{F}=0\rangle$ state it is allowed. In Fig.~\ref{2body} we
show the dipolar relaxation rate coefficients as a function of
magnetic field which were calculated via a coupled-channels
calculation by using recent interaction potentials~\cite{kempen04}.
Except for two peaks which signify Feshbach resonances, the loss
rate coefficients are extremely small, $\sim 3$ orders of magnitude
smaller than the corresponding measured rate coefficients, if the
experimental losses were treated as purely two-body related. For
example, at $880$~G the 2-body loss coefficient is $\sim 5\ast
10^{-17}$~cm$^3$/s which yields, given that the atom density is
$\sim10^{12}$ cm$^{-3}$, a life time of $\sim 20000$~s. As a
comparison, the life time in our dipole trap due to vacuum is less
than $100$~s. As a result, we exclude two-body losses from the
analysis described below and determine that the loss processes in
the region of interest are related to three-body recombination. Note
that while we find this mechanism to be negligible for $^7$Li atoms,
it can be important for heavier alkali atoms. For instance,
$^{133}$Cs experiences large dipolar losses caused by the
second-order spin-orbit interaction \cite{kokkelmans98}.

\subsection{Experimental measurement of three-body recombination loss}

Three-body recombination loss rate near a Feshbach resonance has a
general $a^4$ dependence and in the zero-temperature limit it
diverges at the resonance~\cite{Fedichev96}. Above this general
scaling, universal theory predicts log-periodic oscillations of the
three-body recombination loss rate coefficient ($K_{3}$) due to the
presence of Efimov trimer states. For positive scattering lengths
the oscillations are caused by destructive interference conditions
between two possible decay pathways at certain values of
$a$~\cite{Braaten&Hammer06,EsryNielsen99}. For negative scattering
lengths the loss rate coefficient exhibits a resonance enhancement
each time an Efimov trimer state intersects with the continuum
threshold. Hence, a study of $K_{3}$ across a Feshbach resonance
enables the search for an indirect evidence of Efimov trimer states.

\begin{figure*}[h]
{\centering \resizebox*{0.6\textwidth}{0.29\textheight}
{{\includegraphics{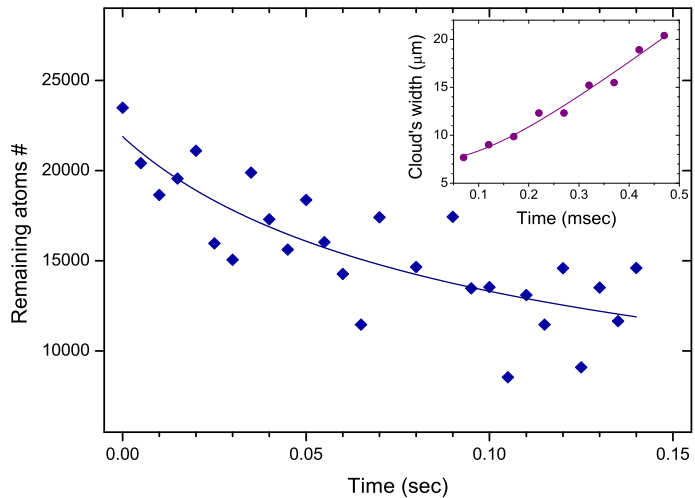}}}
\par}
\caption{\label{LifeTime} A typical atom-number decay measurement on
the $|m_{F}=1\rangle$ state. Inset: time-of-flight measurement from
which the temperature is deduced.}
\end{figure*}

In the experiment, we perform evaporative cooling in an optical
dipole trap near a Feshbach resonance where a gas of $^7$Li atoms is
spontaneously spin purified to the $|m_{F}=0\rangle$ state and is
cooled down to the threshold of degeneracy~\cite{Gross08}. For
measurements in the absolute ground state ($|m_{F}=1\rangle$), a
rapid adiabatic passage by means of a radio-frequency (rf) sweep at
a low magnetic field is used to transfer the atoms from the
$|m_{F}=0\rangle$ state~\cite{Gross10}. We conduct measurements of
atom-number decay and initial temperature as a function of magnetic
field in the vicinity of wide Feshbach resonances on both states. An
example of such set of measurements is shown in Fig.~\ref{LifeTime}.
We use the temperature measurement, together with a precise
characterization of the trap frequencies, to estimate the initial
atom density. Then, the $K_{3}$ value is extracted by fitting the
atom-number decay measurement with the atom-loss rate equation
solution:
\begin{equation}
\dot{N}=-K_{3}\langle n^2\rangle N - \Gamma N, \label{RateEq}
\end{equation}
where $K_{3}$ and $\Gamma$ are the three- and single-body loss rate
coefficients, respectively. $\Gamma$ is determined independently by
measuring a very long decay tail of a low density sample. Note that
this simplified model does not include several important effects,
one of which is the saturation of $K_{3}$ to a maximal value
$K_{max}$ due to finite temperature (unitarity limit) and it can be
represented as \cite{D'Incao04}:
\begin{equation}
K_{max}=c \frac{125 \pi^{2} \hbar^5}{m^3k_{B}^2T^2}, \label{Kmax}
\end{equation}
where $m$ is the atomic mass, $k_{B}$ is the Boltzmann's constant
and $c$ is a numerical constant distinguishing between the two
threshold regimes of the collision energies, $c=1$ for $a>0$ and
$c\approx0.1$ for $a<0$. To address this limitation, measurements
for which $K_{3} \gtrsim 0.1 K_{max}$ are omitted from the analysis
of Efimov features (see Section~\ref{SubS-Universality}).

\begin{figure*}[h]
{\centering \resizebox*{0.6\textwidth}{0.29\textheight}
{{\includegraphics{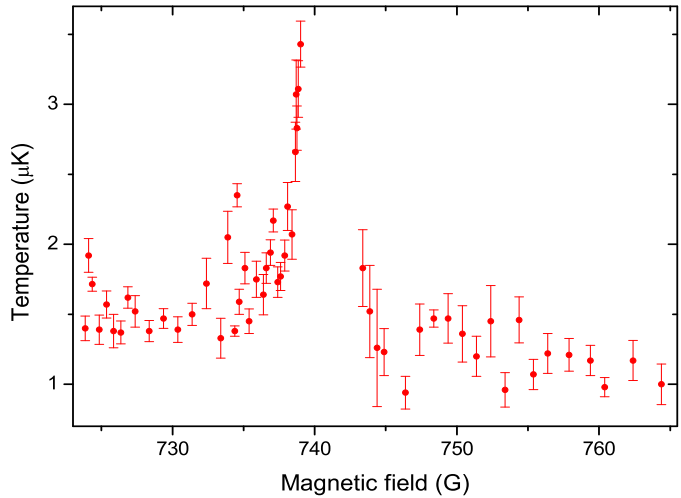}}}
\par}
\caption{\label{Temperature} Initial temperature of the atom-number
decay measurement around the Feshbach resonance on the
$|m_{F}=1\rangle$ state from which the atom density is extracted for
the calculation of $K_{3}$. Each point is deduced from a
time-of-flight measurement (see inset in Fig.~\ref{LifeTime}). The
error bars represent the fitting errors.}
\end{figure*}

Other effects which are not included in the model are
'anti-evaporation' and recombination heating~\cite{Weber03}. The
first is an effective heating caused by a preferential loss of atoms
from the densest (and thus coldest) part of the cloud. We treat the
evolution of our data to no more than $\sim30\%$ decrease in atom
number, as can be seen in Fig.~\ref{LifeTime}, for which
'anti-evaporation' is estimated to induce a systematic error of
$\sim23\%$ towards higher values of $K_{3}$. The error is evaluated
based on an analytical solution to coupled atom-loss rate and
temperature evolution equations~\cite{Weber03,Zaccanti09} and it is
estimated not to limit the accuracy of the reported results.
Recombination heating can be neglected for $K_{3}\ll K_{max}$
because in this case the trap depth (which scales with temperature)
is smaller than the energy released in the recombination process
leading to an immediate loss of the colliding partners. Moreover,
this approach, together with the small decrease in atom number,
allows time evolution of the unitarity limit ($K_{max}(t)$) to be
neglected within a single atom-number decay measurement. Note, that
'anti-evaporation' and recombination heating pose strong limitations
on the measurement's dynamical range in the large scattering lengths
limit where more careful model than Eq.(\ref{RateEq}) has to be
considered. This is demonstrated for $|m_{F}=1\rangle$ state in
Fig.~\ref{Temperature} where a dramatic increase in the initial
temperature of the atom-number decay measurement near the resonance
corresponds to a reduction in the value of $K_{max}$ according to
Eq.(\ref{Kmax}).

\subsection{Efimov features and universality}
\label{SubS-Universality}

\begin{figure*}[h]
{\centering \resizebox*{1\textwidth}{0.3\textheight}
{{\includegraphics{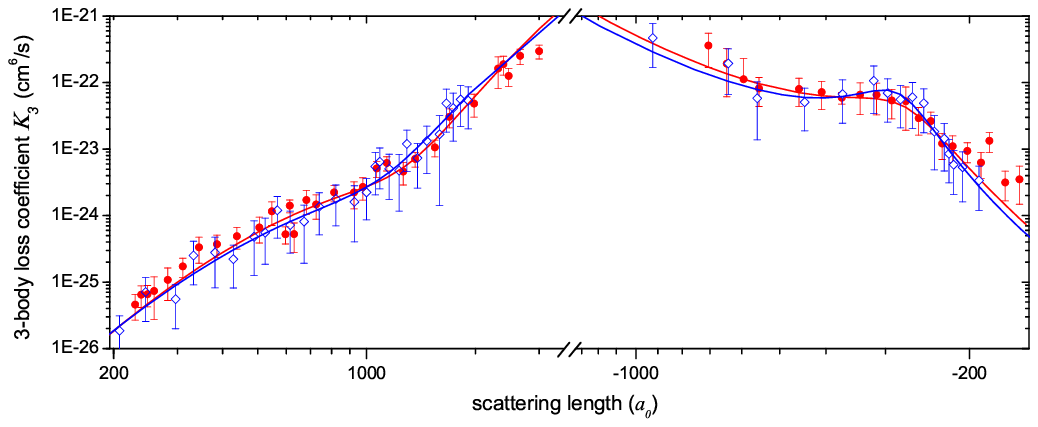}}}
\par}
\caption{\label{K3}Experimentally measured three-body loss
coefficient $K_{3}$ as a function of scattering length (in units
of Bohr radius $a_{0}$) for the $|m_{F}=1\rangle$ state (red solid
circles) and the $|m_{F}=0\rangle$ state (blue open diamonds). The
solid lines (red and blue, respectively) represent fits to the
analytical expressions of universal theory \cite{Kmax}. The error
bars consist of two contributions: the uncertainty in temperature
measurement which affects the estimated atom density and the
fitting error of the atom-number decay measurement.}
\end{figure*}

Experimental results of the three-body loss measurements are
summarized in Fig.~\ref{K3} where $K_{3}$ is plotted as a function
of the scattering length $a$ for the $|m_F=1\rangle$ state (red
solid circles) and the $|m_{F}=0\rangle$ state (blue open
diamonds). Magnetic field values are converted into scattering
lengths by using carefully characterized Feshbach resonances,
discussed in Section~\ref{Sec-FeshbachResonances}. A qualitative
analysis indicates a striking similarity between the two sets of
measurements for both positive and negative scattering lengths. We
can further verify this similarity quantitatively by fitting the
measured $K_{3}$ data to a prediction of universal theory. For
that purpose we represent the loss rate coefficient in a
convenient form~\cite{Braaten&Hammer06}:
\begin{equation}
K_{3}=3C_{\pm}(a)\hbar a^4/m
\end{equation}
where $\pm$ hints at the positive (+) or negative (-) region of the
scattering length. An effective field theory provides analytic
expressions for $C_{\pm}(a)$ that we use in the form represented in
\cite{Kraemer06,Knoop09}:

\begin{equation}
C_{+}(a)=67.1e^{-2\eta_{+}}(\cos^2[s_{0}\ln(a/a_{+})]+\sinh^2\eta_{+})+16.8(1-e^{-4\eta_{+}})
\end{equation}
and
\begin{equation}
C_{-}(a)=4590\sinh(2\eta_{-})/(\sin^2[s_{0}\ln(|a|/a_{-})]+\sinh^2\eta_{-}),
\end{equation}
where the free parameters of the fits are $a_{\pm}$ and $\eta_{\pm}$
which are connected to the real and the imaginary parts of the
three-body parameter,
respectively~\cite{Braaten&Hammer06,Marcelis08}. The fitting results
are represented by solid lines in corresponding colors in
Fig.~\ref{K3}~\cite{Kmax} and the fitting parameters are summarized
in Table \ref{EfimovFeatures}. Comparing corresponding parameters in
different states acknowledges the notion that the Efimov features
are identical within the experimental errors which signifies that
the short-range physics is nuclear-spin independent. An interesting
conclusion can be drawn from this observation. Recall that the short
range potential is given in terms of two-body potential permutations
of the two-body subsystems and a true three-body potential. As the
two Feshbach resonances occur at high magnetic fields where the
nuclear and electron spins are effectively decoupled, the two-body
potentials are similar for both states. Therefore, if the
short-range physics is spin-independent, the true three-body forces
are either also nuclear-spin independent or play a relatively minor
role.

\begin{table}[h]
\begin{center}
\begin{tabular}{p{1.8cm} p{1.5cm} p{1.5cm} p{1.5cm} p{1.5cm} l} \hline\hline
state & $\eta_{+}$ & $\eta_{-}$ & $a_{+}/a_{0}$ & $a_{-}/a_{0}$ & $a_{+}/|a_{-}|$\\
\hline
$|m_{F}=0\rangle$ & $0.213(79)$ & $0.180(48)$ & $238(25)$ & -$280(12)$ & $0.85(11)$ \\

$|m_{F}=1\rangle$ & $0.170(41)$ & $0.253(62)$ & $265(16)$ & -$274(12)$ & $0.97(8)$ \\
\hline\hline
\end{tabular}
\caption{\label{EfimovFeatures}Fitting parameters to universal
theory obtained from the measured $K_{3}$ values of the
$|m_{F}=1\rangle$ and the $|m_{F}=0\rangle$ states.}
\end{center}
\vspace{0cm}
\end{table}

Moreover, two predictions of universal theory are also verified
here: first, the decay parameters $\eta_{+}$ and $\eta_{-}$, which
describe the lifetime of the Efimov state, are assumed to be equal
and indeed they are (within the experimental errors). This suggests
that the imaginary part of the three-body parameter is identical.
Second, the theoretical assumption that the real part of the
three-body parameter across a Feshbach resonance is the same for
negative and positive scattering length regions requires $a_{+}$ and
$a_{-}$ to obey a universal ratio
$a_{+}/|a_{-}|=0.96(3)$~\cite{Braaten&Hammer06}. The fits yield
values of $0.85(11)$ and $0.97(8)$ for $|m_{F}=0\rangle$ and
$|m_{F}=1\rangle$, respectively, which overlap with each other and
with the predicted value within the experimental and theoretical
error bars. We thus confirm that the three-body parameter is
preserved across two different Feshbach resonances and between two
different nuclear-spin states.

For positive scattering lengths, the Efimov trimer is expected to
intersect with the atom-dimer threshold at $a_{*}\approx1.1a_{+}$
\cite{Braaten&Hammer06}. Theory predicts that $a_{*}$ and $a_{-}$
of the same trimer state are related as $a_{-}\approx-22a_{*}$
\cite{Braaten&Hammer06}. Therefore, if the observed Efimov
resonance at $a_{-}$ indicates the lowest state, the one expected
at $a_{*}$ indicates the first excited state as the lowest one
becomes nonuniversal.

\section{Characterization of Feshbach resonances}
\label{Sec-FeshbachResonances}
\subsection{Experimental measurements of binding energy of Feshbach molecules}

\begin{figure*}[h]
{\centering \resizebox*{0.6\textwidth}{0.28\textheight}
{{\includegraphics{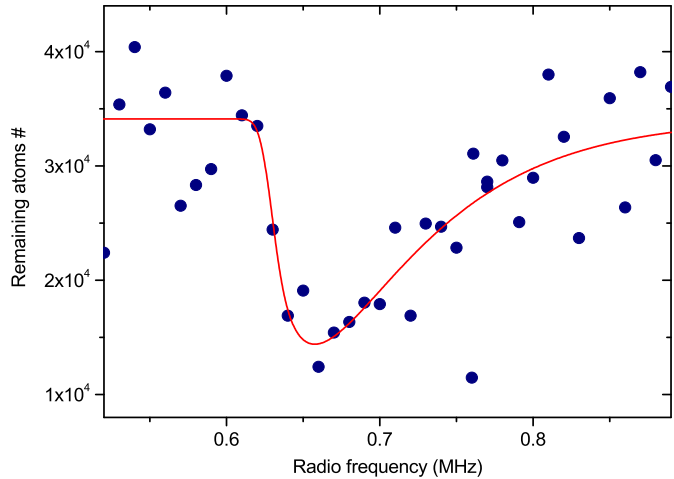}}}
\par}
\caption{\label{RFscan}rf association of molecules at $B=734.4$~G on
the $|m_F=1\rangle$ state. The loss resonance is fitted numerically
to a convolution of Maxwell-Boltzmann and a Gaussian distributions
(solid line).}
\end{figure*}

\begin{figure*}[h]
{\centering \resizebox*{1\textwidth}{0.28\textheight}
{{\includegraphics{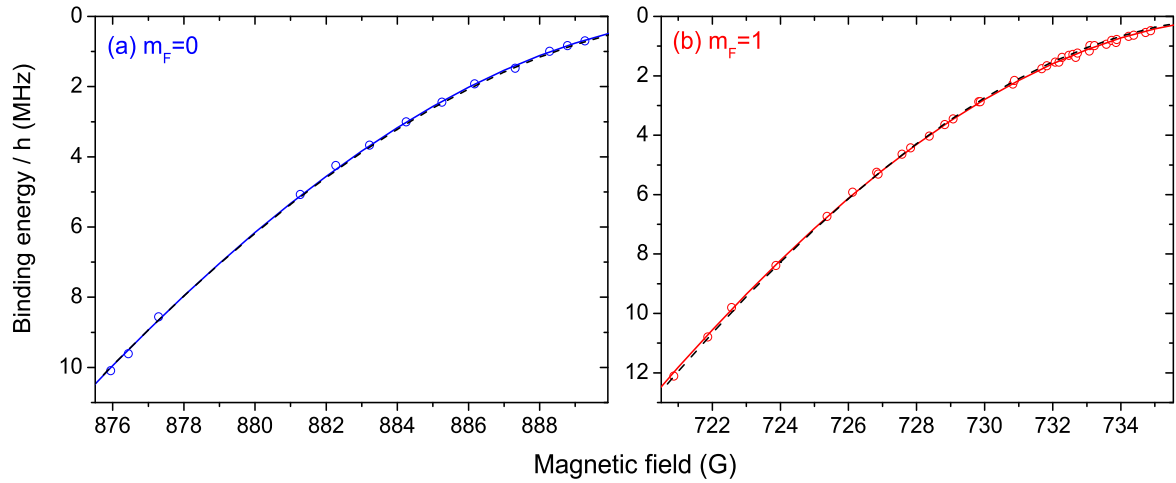}}}
\par}
\caption{\label{BindingE}rf spectroscopy of the molecular binding
energy near the Feshbach resonance in the $|m_{F}=0\rangle$ (a) and
the $|m_{F}=1\rangle$ (b) states. The solid line represents an
independent fit to CC calculation while the dark dashed line is the
combined fit for both states simultaneously.}
\end{figure*}

For a correct investigation of the $K_3$ coefficients dependence on
the two-body scattering length, it is crucial to have an accurate
mapping between the scattering length and the applied magnetic
field. For this purpose, the binding energies of the underlying
bound states which give rise to the two broad Feshbach resonances
are carefully measured using rf molecule association. This method
uses a weak rf field to resonantly associate weakly bound Feshbach
dimers which are then rapidly lost through collisional relaxation
into deeply bound states~\cite{Thompson05}. In the experiment the rf
modulation time is varied between 0.5 and 3 sec and the modulation
amplitude ranges from 150 to 750~mG. The remaining atom number is
measured by absorption imaging as a function of rf frequency at a
given magnetic field and the rf-induced losses are then numerically
fitted to a convolution of a Maxwell-Boltzmann and a Gaussian
distributions. The former accounts for broadening of the
spectroscopic feature due to finite kinetic energy of atoms at a
typical temperature of $\sim 1.5\mu K$~\cite{Hanna07}. The latter
reflects broadening due to magnetic field instability and
shot-to-shot atom number fluctuations. From the fit we extract the
molecular binding energy ($E_{b}$) corresponding to zero
temperature. An example of a molecule association induced loss
feature is depicted in Fig.~\ref{RFscan} where the characteristic
asymmetry of the obtained profile is clearly seen~\cite{Hanna07}.
Results of the binding energy rf spectroscopy for both states are
shown in Fig.~\ref{BindingE}. We analyze these results by fitting
them with a coupled-channels calculation which is discussed in the
next Section.

\subsection{Analysis of two-body interactions}

Very close to resonance, the binding energy is related to the
scattering length as

\begin{equation}
 E_b(a)=-\frac{\hbar^2}{m a^2},
\end{equation}
This expression shows that ultracold scattering physics is closely
related to bound state physics just below the dissociation
threshold. However, this simple relationship rapidly breaks down
when going further away from resonance, and is therefore not
sufficient to interpret three-body recombination in terms of the
scattering length.

In order to make a correct mapping between binding energy and
scattering length, we make use of a coupled-channels two-body
interaction model for lithium, which has been discussed
earlier~\cite{kempen04}, and we use the interaction potentials
discussed there as a starting point for our analysis.  For large
interatomic separations $r$ the singlet ($S=0$) and a triplet
($S=1$) potentials are given by

\begin{equation}
 V_S(r)=-\frac{C_6}{r^6} -\frac{C_8}{r^8}-\frac{C_{10}}{r^{10}} -(-1)^{S} C^{ex} r^{7/2\alpha -1}e^{-2 \alpha r},
\end{equation}
with van der Waals coefficients $C_6, C_8, C_{10}$~\cite{Yan96},
exchange parameter $C^{ex}$~\cite{Zemke99,Marinescu96}, and the
ionization energy $\alpha ^2/2$~\cite{NISTwww}. For the short
radial range $r$ we use model singlet and triplet potentials which
have also been used in Refs.~\cite{Moerdijk94,vanAbeelen97}.

The short-range and long-range potentials are smoothly connected at
$r=18a_0$. We improve the accuracies of the short range potentials
considerably, by making use of the accumulated phase
method~\cite{Moerdijk94}. Therefore a boundary condition is applied
on the partial-wave radial wave functions at $r=7a_0$ in the form of
a WKB phase

\begin{equation}
 \phi_{S,T}(E,\ell)=\phi^0_{S,T}(E,\ell)+\Delta \phi_{S,T}.
\end{equation}

The first term on the right is calculated by radial integration of
the model potential up to $7 a_0$ and is sufficiently accurate to
account for the energy and angular momentum dependence of the
accumulated phase. The corrections $\Delta \phi_{S,T}$ to the
accumulated singlet and triplet phases are independent of energy
and angular momentum.

The most crucial parameters in the coupled channels model are
$\Delta \phi_{S}$, $\Delta \phi_{T}$ and $C_6$, which we take as
free parameters that we determine from our experimental binding
energies. We perform a $\chi ^2$ minimization with respect to the
free parameters, and this allows us to determine the positions of
both Feshbach resonances, and the direct mapping of the scattering
length on the magnetic field $a(B)$. In order to extract the
resonance's parameters we fit $a(B)$ with a factorized
expression~\cite{Lange09}:

\begin{equation}
\label{ScatteringLength}
\frac{a}{a_{bg}}=\prod_{i=1}^{N}\\
{\left(1-\frac{\Delta_{i}}{(B-B_{0,i})}\right)},
\end{equation}

where $a_{bg}$ is the background scattering length, $\Delta_{i}$ is
the $i$'s resonance width and $B_{0,i}$ is the $i$'s resonance
position.

We analyze the data in two different ways. First we perform an
independent fit of the model to the $|m_{F}=0\rangle$ and the
$|m_{F}=1\rangle$ binding energies, represented by solid blue and
red lines in Fig.~\ref{BindingE}(a) and (b), respectively. Then, a
combined fit for both $m_{F}$ states is performed (dashed black line
in Fig.~\ref{BindingE}(a,b)), which allows us to check the
consistency between the two different experiments, within the same
interaction model. Deviations of the second case from the first one
are a hint for possible different systematic shifts in magnetic
field calibration for both states. This can be explained by the fact
that a precise calibration of the magnetic field is performed
locally by microwave transitions in the vicinity of each of the wide
resonances. Note that each set of measurements is limited to a range
of $\pm 20$~G around the resonances while the distance between them
is $\sim 150$~G. We therefore use the results of the independent
fits for the $K_{3}$ analysis which was discussed previously in this
paper. The resonances' parameters as obtained from the independent
fits are summarized in Table~\ref{B0individual}.

\begin{table}[h]
\begin{center}
\begin{tabular}{p{1.8cm} p{1.5cm} p{1.5cm} p{1.5cm} l}
\hline\hline
state & type & $a_{bg}/a_{0}$ & $\Delta$~(G) & $B_{0}$~(G) \\
\hline
$|m_{F}=0\rangle$& narrow & $-18.24$ & $+4.518$ & $846.0$ \\

$|m_{F}=0\rangle$& wide & $-18.24$ & $-237.8$ & $893.7(3)$ \\

$|m_{F}=1\rangle$& wide & $-20.98$ & $-171.0$ & $738.2(2)$ \\
\hline\hline
\end{tabular}
\caption{\label{B0individual} Feshbach resonance parameters for
both states as obtained from independent fits of the CC
calculation to the molecular binding energies (solid lines in
Fig.~\ref{BindingE}).}
\end{center}
\vspace{0cm}
\end{table}

% \begin{eqnarray}
%  B_0^{m_f=1}=738.20(0.18) G, \\
%  B_0^{m_f=0}=893.74(0.29) G, \\
% \end{eqnarray}
% and for the combined fit
% \begin{eqnarray}
%  B_0^{m_f=1}=737.91(0.02) G, \\
%  B_0^{m_f=0}=893.87(0.05) G. \\
% \end{eqnarray}

Table \ref{B0combined} represents the resonances' positions obtained
from the combined fit and compares them with our most precise
to-date experimental values, i.e. an atom loss measurement for the
narrow resonance of $|m_{F}=0\rangle$~\cite{Gross09} and the two
independent fits of the wide resonances (Table~\ref{B0individual}).
A field calibration uncertainty of $0.3$~G is added to the fitting
errors of the experimental values.

\begin{table}[h]
\begin{center}
\begin{tabular}{p{1.8cm} p{2cm} p{2.4cm} l}
\hline\hline
state & type & \multicolumn{2}{c}{$B_{0}$~(G)} \\
 & & Combined fit & Experimental\\
\hline
$|m_{F}=0\rangle$& narrow & $845.54$ & $844.9(8)$\\

$|m_{F}=0\rangle$& wide & $893.95(5)$ & $893.7(4)$\\

$|m_{F}=1\rangle$& wide & $737.88(2)$ & $738.2(4)$\\
\hline\hline
\end{tabular}
\caption{\label{B0combined}Feshbach resonances' positions ($B_{0}$)
as obtained from a combined fitting of the molecular binding energy
measurements in both states simultaneously to the CC calculation.
The experimentally determined positions are presented in the last
column where the narrow resonance was determined by atom loss
measurement \cite{Gross09} and the two wide ones where determined by
independent fits of the molecular binding energies (see Table
\ref{B0individual}). The field calibration uncertainty is included
in the experimental errors.}
\end{center}
\vspace{0cm}
\end{table}

We find the results in Table \ref{B0combined} mutually consistent.
Moreover, the determined values for $\phi_{S}$, $\Delta \phi_{T}$
and $C_6$ are all consistent with the bound of earlier performed
analysis~\cite{Moerdijk94,Abraham97,vanAbeelen97,kempen04}. Since
our combined analysis only weakly depends on the value of the $C_6$
coefficient, we present our results with a fixed value of $C_6$
which was found in the {\it ab initio}
calculations~\cite{Yan96,Derevianko01}. The determined parameters
$\phi_{S}$ and $\Delta \phi_{T}$ correspond to the singlet and
triplet scattering lengths $a_{S}$=34.33(2) $a_0$ and
$a_{T}$=-26.87(8) $a_0$.

\section{Conclusions}
\label{Sec-Conclusions}

In this work we study experimentally Efimov physics in two
nuclear-spin sublevels of bosonic lithium and show that the
positions and widths of recombination minima and Efimov resonances
are identical for both states within the experimental errors. As the
properties of Efimov features are governed by the three-body
parameter, our study indicates that the short-range physics is
nuclear-spin independent. We also find that the Efimov features are
universally related across the Feshbach resonances. We note that
slight deviations of our measurements from the universally predicted
values can be explained by finite effective range corrections which
were recently evaluated by means of an effective field
theory~\cite{Ji10}. Let us note also that the observed position of
the Efimov resonance reveals the same numerical factor $|a_{-}|/r_0
\approx 8.5$ as in the experiments on $^{133}$Cs \cite{Kraemer06}
which may or may not be an accidental coincidence.

The reported results crucially depend on a careful mapping between
the scattering length and the magnetic field. We characterize two
wide Feshbach resonances in different states by fitting the binding
energies of weakly bound molecules, created by rf-association, with
a coupled channels analysis. This gives rise to a very precise
determination of the absolute positions of the Feshbach resonances
and the values of the singlet and triplet scattering length that
characterize the molecular potentials of lithium.

\section{Acknowledgments}
This work was supported, in part, by the Israel Science Foundation
and by the Netherlands Organization for Scientific Research (NWO).
N.G. is supported by the Adams Fellowship Program of the Israel
Academy of Sciences and Humanities.

\label{}
% etc, etc

% The Appendices part is started with the command \appendix;
% appendix sections are then done as normal sections
% \appendix

% \section{}
% \label{}

% The Acknowledgements are also a un-numbered section
%\section*{Acknowledgements}
% Acknowledgements text here


\begin{thebibliography}{00}
\bibitem{Kmax} We note that the four points with the highest values of $K_{3}$ on $|m_{F}=1\rangle$ $a>0$ were omitted from the
fitting procedure on the grounds of approching the finite
temperature saturation value $K_{max}$ (unitarity limit)
\cite{D'Incao04} which is $\sim4.5\times10^{-21}$ cm$^{6}$/s for a
temperature of $3.5$~$\mu K$.
\end{thebibliography}
\end{document}